\documentclass[runningheads]{llncs}
\usepackage[T1]{fontenc}
\usepackage{graphicx}
\usepackage{amsmath}
\usepackage[nolist,nohyperlinks]{acronym}
\usepackage{amssymb}
\usepackage{comment}
\usepackage{marvosym}

\usepackage{hyperref}
\hypersetup{colorlinks}

\acrodef{CAS}{Computer-Assisted Surgery}
\acrodef{MIS}{Minimally Invasive Surgery}
\acrodef{SOTA}{state-of-the-art}
\acrodef{STTN}{Spatial-Temporal Transformer Network}
\acrodef{DAEVI}{Depth-Aware Endoscopic Video Inpainting}
\acrodef{STGDE}{Spatial-Temporal Guided Depth Estimation}
\acrodef{BMPCF}{Bi-Modal Paired Channel Fusion}
\acrodef{DED}{Depth-Enhanced Discriminator}
\acrodef{GAN}{Generative Adversarial Network}
\acrodef{PSNR}{Peak Signal-to-Noise Ratio}
\acrodef{MSE}{Mean Squared Error}
\begin{document}
\title{Depth-Aware Endoscopic Video Inpainting}
%
%
\author{Francis Xiatian Zhang\inst{1}\orcidID{0000-0003-0228-6359} \and
Shuang Chen\inst{1}\orcidID{0000-0002-6879-7285} \and
Xianghua Xie\inst{2}\orcidID{0000-0002-2701-8660} \and\\
Hubert P. H. Shum \inst{1}\textsuperscript{(\Letter)} \orcidID{0000-0001-5651-6039}}
%
\authorrunning{Zhang et al.}
%
\institute{
Department of Computer Science, Durham University, 
Durham, United Kingdom\\
\email{\{xiatian.zhang, shuang.chen, hubert.shum\}@durham.ac.uk}
\and
Department of Computer Science, Swansea University, Swansea, United Kingdom\\
\email{x.xie@swansea.ac.uk}
}
\maketitle              

\begin{abstract}
Video inpainting fills in corrupted video content with plausible replacements. While recent advances in endoscopic video inpainting have shown potential for enhancing the quality of endoscopic videos, they mainly repair 2D visual information without effectively preserving crucial 3D spatial details for clinical reference. Depth-aware inpainting methods attempt to preserve these details by incorporating depth information. Still, in endoscopic contexts, they face challenges including reliance on pre-acquired depth maps, less effective fusion designs, and ignorance of the fidelity of 3D spatial details.
To address them, we introduce a novel Depth-aware Endoscopic Video Inpainting (DAEVI) framework. It features a Spatial-Temporal Guided Depth Estimation module for direct depth estimation from visual features, a Bi-Modal Paired Channel Fusion module for effective channel-by-channel fusion of visual and depth information, and a Depth Enhanced Discriminator to assess the fidelity of the RGB-D sequence comprised of the inpainted frames and estimated depth images.
Experimental evaluations on established benchmarks demonstrate our framework's superiority, achieving a 2\% improvement in PSNR and a 6\% reduction in MSE compared to state-of-the-art methods. Qualitative analyses further validate its enhanced ability to inpaint fine details, highlighting the benefits of integrating depth information into endoscopic inpainting.
\end{abstract}
\section{Introduction}

In endoscopic videos, occlusions or artifacts caused by reflections or instrument shadows significantly degrade the visual quality. This issue is commonly known as \emph{corruptions}, hiding critical anatomical details required for endoscopy examinations and surgeries, affecting clinical decision significantly \cite{monkam2021easyspec}.  
As a technique to improve video quality by reconstructing the corrupted regions based on the uncorrupted information, video inpainting is introduced by \cite{shen2020content,tukra2021see,ali2021deep,daher2023temporal} into the endoscopic scenario to mitigate the corruptions, known as endoscopic video inpainting. While these methods could reconstruct 2D visual information in corrupted endoscopic videos, they suffer from preserving vital 3D spatial details, resulting in the artifact and spatial inconsistency at the inpainted regions, such low-fidelity performance limits their reliability for clinical applications. 

Employing depth map to complement 3D spatial information is wildly applied in general video painting~\cite{liao2020dvi,yamashita2022boundary,li2023depth}, which offers a promising solution to preserve 3D spatial awareness for endoscopic video inpainting. Nevertheless, applying this solution is hindered by three significant challenges: First, it is not feasible to pre-acquire endoscopic depth maps, as the depth sensor is not available in standard monocular endoscopic cameras~\cite{fu2021future}. Second, given the learned depth features from deep-learning-based methods, simply concatenating visual features channel-wise and using vanilla convolution for fusion~\cite{li2023depth} fails to effectively exploit the depth representation, as they tend to capture redundant representations from visual features~\cite{li2023scconv}, losing the 3D spatial details complementing by depth maps. Third, none of these methods~\cite{liao2020dvi,yamashita2022boundary,li2023depth} assess the 3D spatial fidelity in RGB inpainted outputs, which compromises the reliability of inpainted content.

To address these challenges, we propose the \ac{DAEVI} framework that provides more reliable inpainted details for improved clinical reference.
It consists of a \ac{STGDE} module, a \ac{BMPCF} module, and a \ac{DED}, each designed to overcome the respective challenge.
First, our STGDE module extracts depth information during visual feature learning to provide 3D spatial information, thus avoiding the requirement for pre-acquired depth maps as input.
Second, the BMPCF module conducts a tailor-made feature fusion algorithm to better correlate the 3D spatial relevancy between visual and depth features by pair-wise fusing each visual and depth feature.
Third, our DED assesses the 3D spatial fidelity of the RGB-D sequence formed by the inpainted frames and estimated depths, promoting realistic outputs with plausible 3D spatial details.
\begin{figure}[t]
    \centering
    \includegraphics[scale=0.31]{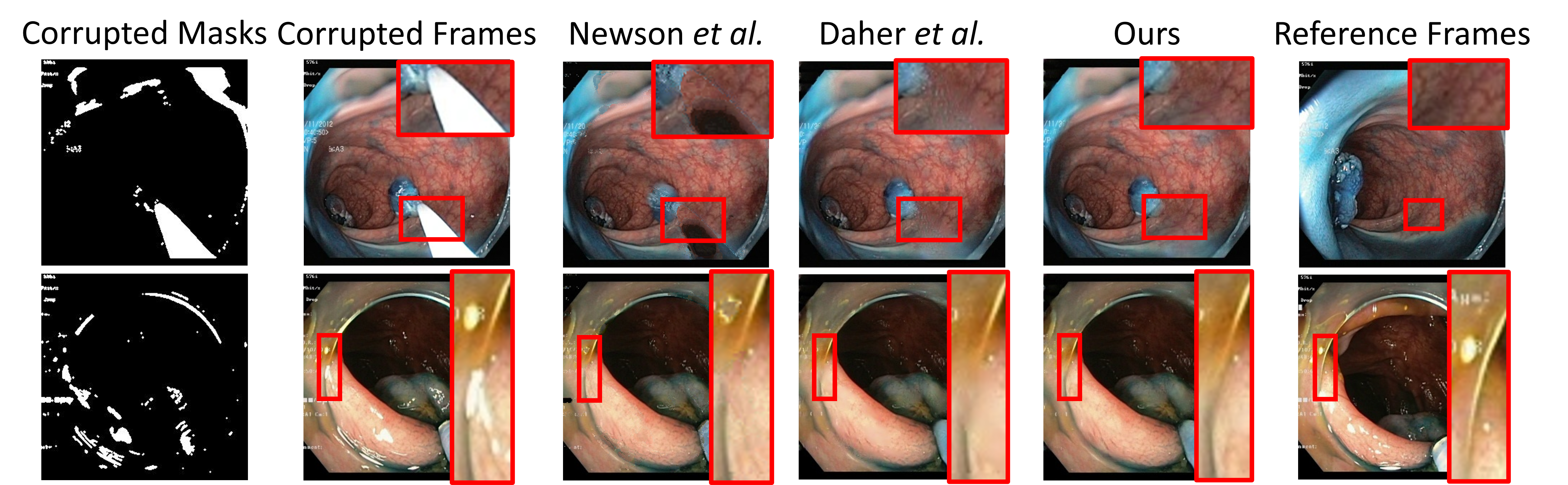}
    \caption{Comparison with previous methods by Newson \emph{et al.} \cite{newson2014video} and Daher \emph{et al.} \cite{daher2023temporal} on corrupted frames from the HyperKvasir dataset \cite{borgli2020hyperkvasir}. Red boxes highlight significant differences. Reference frames are near frames with less corruption. Our inpainted content is not only visually plausible but also contextually realistic.}
    \label{fig:demo}
\end{figure}

We evaluate our method on the HyperKvasir endoscopic video dataset~\cite{borgli2020hyperkvasir} and compare it with the corresponding benchmark~\cite{zeng2020learning,daher2023temporal}. The quantitative experiments demonstrate that our proposed DAEVI outperforms state-of-the-art approaches~\cite{daher2023temporal}, achieving approximately 2\% better \ac{PSNR} and 6\% lower \ac{MSE}. Our qualitative results show that DAEVI inpaints more fine-grained details, such as microvessels and the boundary of instruments, as depicted in Fig.~\ref{fig:demo}. Furthermore, we directly apply our DAEVI trained on HyperKvasir to the SERV-CT datasets~\cite{edwards2022serv}, demonstrating our method's generalizability in endoscopic video inpainting. Our source code is available on \textcolor{blue}{\url{https://github.com/FrancisXZhang/DAEVI}}.

Our work contributes in several ways, as outlined below:
\begin{enumerate}
\item To the best of our knowledge, \ac{DAEVI} is the first endoscopic video inpainting framework to incorporate depth information. The effectiveness is demonstrated by comprehensive experiments.
\item We propose a Spatial-Temporal Guided Depth Estimation module, to translate depth representation directly from latent visual features, hence circumventing the challenge of acquiring depth maps during endoscopic surgery.
\item We design a  Bi-Modal Paired Channel Fusion module that fuses each pair of channels from visual and depth features, which effectively leverages 3D spatial details in endoscopic inpainting.
\item We introduce a Depth-Enhanced Discriminator within our end-to-end optimization, which assesses the fidelity of the inpainted RGB-D sequence, promoting realistic outputs with more plausible 3D spatial details.
\end{enumerate}

\section{Methodology}
Given the input endoscopic video frames $X \in \mathbb{R}^{T \times H \times W \times 3}$, we leverage the binary mask $M \in \mathbb{R}^{T \times H \times W \times 1}$, which identifies the corrupted regions, to get the input  $X_{M} = {X}\odot{M}$. After processing by our \ac{DAEVI} to generate the uncorrupted output $\hat{Y} \in \mathbb{R}^{T \times H \times W \times 3}$. $\odot$ is the element-wise product, $H\times W$ is the spatial dimension. 
The whole formulation is denoted as: ${\hat{Y}}=DAEVI({X}_{M})$.

\begin{figure}[t]
    \centering
    \includegraphics[scale = 0.39]{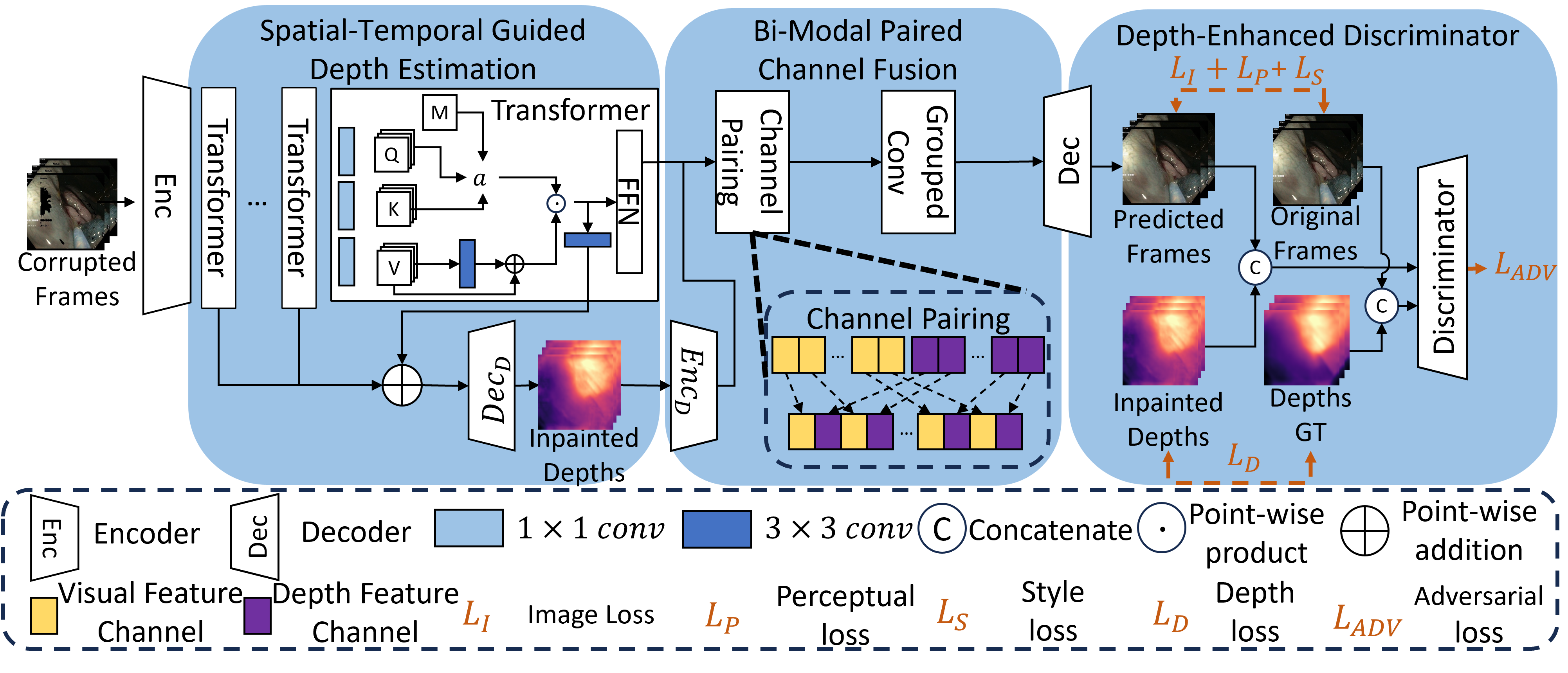}
    \caption{The overview of our framework. First, our Spatial-Temporal Guided Depth Estimation module translates depth information from corrupted frames  (See \ref{sec:stgde}). Second, our Bi-Modal Paired Channel Fusion module effectively fuses visual features with depth features  (See \ref{sec:BMPCF}). Third, our Depth Enhanced Discriminator assesses the fidelity of the inpainted RGB-D sequence  (See \ref{sec:DED}).}
    \label{fig:overview}
\end{figure}
The overall architecture is shown in Fig.~\ref{fig:overview}. 
First, DAEVI employs a convolutional encoder to embed $X_M$ into compact visual features $F \in \mathbb{R}^{\frac{H}{4}\times \frac{W}{4}\times 4C}$
to effectively represent local visual features.
Following this, our Spatial-Temporal Guided Depth Estimation (STGDE) module learns multi-level visual features and translates them into depth maps.
Subsequently, the proposed Bi-Modal Paired Channel Fusion (BMPCF) module fuses visual and depth features to obtain an integrated representation with enhanced 3D spatial details.
After that, a convolutional decoder reconstructs the final inpainted frames $\hat{Y}$. 
During training, our Depth-Enhanced Discriminator (DED) assesses the visual and spatial fidelities of the inpainted RGB-D sequence.
\subsection{Spatial-Temporal Guided Depth Estimation (STGDE)}
\label{sec:stgde}
Depth-aware endoscopic video inpainting faces a unique challenge in acquiring depth data, as the standard endoscopic cameras are technically unable to provide the raw depth~\cite{fu2021future}. 
To address this challenge, we propose a \ac{STGDE} module to translate depth features from latent visual features. STGDE involves multiple transformer blocks $TB$ and a depth decoder $Dec_D$. Each TB is a spatial-temporal transformer block~\cite{zeng2020learning} with a spatial enhancement to enhance the local feature learning, thereby improving the representation ability. $Dec_D$ aims to gather the latent visual feature across all $TBs$ for effective depth estimation.

Specifically, after the encoder, corrupted frames $X_M$ are embedded as visual features $F$, which are fed into the first transformer block $TB_1$. The output from $TB_i$ is subsequently fed into the following $TB_{i+1}$, where $i \in (1,{N_s}-1)$ and ${N_s}$ is the number of $TB_s$. The $F^{i-1}$, as the input for each $TB_i$, are linearly transformed into query $Q^i$, key $K^i$, and value $V^i$ separately. We introduce a $3\times3$ depth-wise convolution $P'^i_{V}(\cdot)$ to enhance the $V^i$ for better spatial feature learning:
\begin{equation}
Q^i, K^i, V^i = P^i_{Q}(F^{i-1}), P^i_{K}(F^{i-1}), P^i_{V}(F^{i-1}) + P'^i_{V}(F^{i-1}),
\label{eq:embed}
\end{equation}
where $P^i_{Q,K,V}(\cdot)$ denote 1x1 2D convolutions. After that, we split each of $Q^i$, $K^i$, and $V^i$ into $n={r_1}\times {r_2}$ smaller patches $Q^i_p$, $K^i_p$, and $V^i_p$ $\in \mathbb{R}^{Tn \times c \times h/r_1 \times w/r_2}$, where $h/r_1 \times w/r_2$ is the spatial dimension of patches. Then we utilize the patched $Q^i_p$, $K^i_p$, and $V^i_p$ to get the attention output $F^{i}_{att}$:
\begin{equation}
S^i = Q^{i}_p(K^{i}_p)^\top/\sqrt{r_1 \times r_2 \times c},
\label{eq:score}
\end{equation}
\begin{equation}
F^{i}_{att} = \text{softmax}(S^i \odot M)V^i_p,
\label{eq:attn}
\end{equation}
where $S^i$ is the attention score, and $M$ denotes a resized binary mask matrix indicating the currupted regions~\cite{zeng2020learning}. Then, after a convolutional projection $P^i_F$ followed by a feed-forward network $FFN^{i}$, we obtain the output $F^{i}$ of $TB_i$:
\begin{equation}
F^{i} = FFN^i\left(P^i_F\left(F^{i}_{att}\right)\right).
\end{equation}

To translate depth maps $\hat{D}$ from latent visual features, we aggregate $F_{att}$ across all $TB_s$ to gather the multi-layer visual representation. After a convolutional projection $P^i_D$, the depth decoder generate the depth maps $\hat{D}$:
\begin{equation}
\hat{D} = Dec_D\left(\sum_{i=1}^{N_s} P^i_D\left(F^{i}_{att}\right)\right).
\end{equation}

\subsection{Bi-Modal Paired Channel Fusion (BMPCF)}
\label{sec:BMPCF}
Endoscopic depth-aware inpainting encounters a challenge in effectively integrating depth with visual information, as the simple channel-wise concatenation followed by a vanilla convolution is unable to fully exploit the correlation between depth and visual feature, especially in endoscopic scenes such complex nature involving varied spatial structures~\cite{shao2022self}.

To address this challenge and effectively enhance visual information with 3D spatial details, we design a \ac{BMPCF} module to correlate each visual and depth feature by a tailor-made pair-wise fusion algorithm. 

Specifically, given the depth maps $\hat{D}$ translated by $Dec_D$, we first enlarge the channel capacity of $\hat{D}$ with a depth encoder $Enc_D$ to obtain the embedded depth feature $F_D$, which has the same number of channels as the STGDE's output $F^{N_{s}}$. Then, to ensure each depth correlates precisely to the corresponding visual feature, we sequentially interleave the sliced $F^{N_{s}}$ and $F_D$ in channel-wise:

\begin{equation}
F_{pair}[:, 2i] =
\begin{cases}
F^{N{s}}[:, i] & \text{for } i = 0, 2, \ldots, c-2, \\
F_D[:, i] & \text{for } i = 1, 3, \ldots, c-1,
\end{cases}
\end{equation}
where $c$ is the channel number of $F^{N{s}}$ and $F_D$, also indicates the number of pairs. After that, the group-wise convolution $G(\cdot)$ \cite{krizhevsky2012imagenet} is employed on $F_{pair}$ to fuse each pair of visual and depth features, producing the fused output $F_{f} \in \mathbb{R}^{T \times c \times h \times w}$:
\begin{equation}
F_{f} = G(F_{pair}).
\end{equation}

In this way, each convolutional kernel facilitates the fusion between every two adjacent channels consisting of one visual feature and one depth feature. Subsequently, a convolutional decoder reconstructs inpainted frames $\hat{Y}$ from $F_{f}$.

\subsection{Depth-Enhanced Discriminator (DED)}
\label{sec:DED}
Effectively assessing the spatial fidelity is critical in endoscopic depth-aware inpainting, as minor inaccuracies, such as incorrect anatomical details, could significantly impact clinical decisions \cite{dray2023artificial}. While \ac{GAN} strategies \cite{goodfellow2020generative} in previous depth-aware inpainting methods \cite{yamashita2022boundary,li2023depth} only enhance the fidelity of RGB content and ignore the fidelity in 3D spatial details, resulting in unreliable outputs for clinical reference \cite{li2023depth}.
To this end, we introduce \ac{DED} during training, to comprehensively assess the RGB-D inpainted endoscopic frames across spatial, temporal and depth dimensions.

Specifically, we followed \cite{chang2019free} to build our \ac{DED} with 6 convolutional blocks. The inpainted frames and depth are concatenated as the RGB-D data to be the input of DED. The adversarial loss $L_{ADV} = L_{GEN} + L_{DED} $ is adopted from~\cite{goodfellow2020generative}:
\begin{equation}
\begin{split}
L_{DED} = \mathbb{E}_{(D, Y) \sim P_{Data}}[Relu(1 - DED([D, Y]))]  \\ \mathbb{E}_{(\hat{D}, \hat{Y}) \sim P_{G}}[Relu(DED([\hat{D}, \hat{Y}]))],
\end{split}
\end{equation}
\begin{equation}
L_{GEN} = - \mathbb{E}_{(\hat{D}, \hat{Y}) \sim P_{G}}[DED([\hat{D}, \hat{Y}])],
\end{equation}

To enhance parameter optimization efficiency, we adopt an end-to-end optimization strategy for our \ac{DAEVI}. Our full loss function is as follows:
\begin{equation}
L = \lambda_{D}L_{D} + \lambda_{I}L_{I} + \lambda_{GEN}L_{GEN} + \lambda_{P}L_{P} + \lambda_{S}L_{S},
\end{equation}
where $\lambda$ denotes the weight of each loss term. $L_{D}$ and $L_{I}$ are the L1 reconstruction loss~\cite{zhou2023propainter} for translated depth and inpainted frames, respectively. $L_{P}$ and $L_{S}$ denote the perceptual loss and style loss, respectively \cite{johnson2016perceptual} (details can be found in our supplementary material). In each iteration, $L$ and $L_{DED}$ optimize our inpainting network and our \ac{DED}, respectively.

\section{Experiment}
\subsection{Experimental Setting}
We evaluate our method against the existing benchmark established \cite{daher2023temporal} on the HyperKvasir endoscopic video dataset \cite{borgli2020hyperkvasir}. This dataset includes 373 videos with a total of 889,372 frames, 343 for training and 30 for testing. Depth ground truth is derived from a pre-trained endoscopic depth estimator \cite{shao2022self} on unmasked frames, which is needed only in training. Following the benchmark setting \cite{daher2023temporal}, we employ their provided masks to identify corrupted regions in frames and calculate metrics such as Peak Signal-to-Noise Ratio (PSNR), Structural Similarity Index (SSIM), and Mean Square Error (MSE) specifically for corrupted regions.

We configure the block number $N_s = 8$ and set the weights for $\lambda_{D}$, $\lambda_{P}$, $\lambda_{S}$, $\lambda_{I}$, and $\lambda_{GEN}$ to [0.1, 0.1, 250, 1, 0.01]. We employ Adam optimizer with learning rate $=$ 1e-4, $\beta_1$: 0, $\beta_2$: 0.99. All experiments are trained on an NVIDIA TITAN RTX 24G GPU with a batch size of 4 for 200k iterations. Training iterations alternate between selecting 5 random or consecutive frames resized to 288 x 288 pixels. For inference, the model processes every 5 corrupted frames alongside 10 nearby corrupted frames sampled for reference, reassembling them into the full video with a real-time processing speed of approximately 0.03 seconds per frame.

\begin{table}[t]
\scriptsize
\centering
\caption{Inpainting Performance Comparison and Ablation Study. w/o STGDE: A pre-trained depth estimator \cite{shao2022self} is leveraged for depth estimation instead of \ac{STGDE}; w/o BMPCF: Simple concatenation is used for fusion instead of \ac{BMPCF}; w/o DED: A standard RGB discriminator \cite{chang2019free} is used in \ac{GAN} training instead of \ac{DED}.}
\begin{tabular}{l|p{2.3cm}|p{2.3cm}|p{2.3cm}}
\hline
Methods & $PSNR_{Crop}\uparrow$  & $SSIM_{Crop} \uparrow$   &  $MSE_{Crop} \downarrow$  \\
\hline
Arnold \emph{et al.} \cite{arnold2010automatic} & 19.909 & 0.559 &  895.222 \\
Newson \emph{et al.} \cite{newson2014video} & 22.27 & 0.650 &  543.636 \\
STTN \cite{zeng2020learning} & 28.683 & \underline{0.793} &  119.541 \\
Daher \emph{et al.} \cite{daher2023temporal} & 29.542 & 0.785 &  104.719 \\
\hline
DAEVI (Full Framework) & \textbf{30.126} & \textbf{0.797} &  \textbf{97.873}\\
w/o \ac{STGDE} & \underline{29.801} & 0.788 & 105.150\\
w/o \ac{BMPCF} &29.695&0.791&\underline{103.861}\\
w/o \ac{DED} & 29.286 & \textbf{0.797} & 108.903\\
\hline
\end{tabular}
\label{tab:SOTA}
\end{table}

\subsection{Results}
\noindent \textbf{Comparison with Existing Methods} In Table \ref{tab:SOTA}, we benchmark our \ac{DAEVI} against Arnold \emph{et al.} \cite{arnold2010automatic}, a diffusion-based approach; Newson \emph{et al.} \cite{newson2014video}, employing temporal patches; STTN \cite{zeng2020learning}, a transformer-based method; and Daher \emph{et al.} \cite{daher2023temporal}, the \ac{SOTA} method in endoscopic video inpainting. Our method outperforms existing methods across all metrics. These results highlight the benefits of integrating depth information into endoscopic video inpainting. Clinically, these improvements enhance the visibility of endoscopic videos \cite{tukra2021see}, crucial for accurate diagnostics and surgical planning.

\noindent \textbf{Qualitative Results}
To demonstrate our inpainting performance on corrupted regions, we remove specular reflections and high reflection from instruments in the HyperKvasir dataset \cite{borgli2020hyperkvasir} and the SERV-CT dataset, an external-body endoscopic dataset with depth ground truth provided \cite{edwards2022serv}. Figure \ref{fig:demo} from the HyperKvasir dataset illustrates our method more effectively restoring details such as microvessels and the interface between instruments and organs. These enhancements are crucial for safer and more efficient endoscopic operations \cite{ben2012adverse}. More qualitative results can be found in our supplementary material.

\begin{figure}[t]
    \centering
    \includegraphics[scale = 0.34]{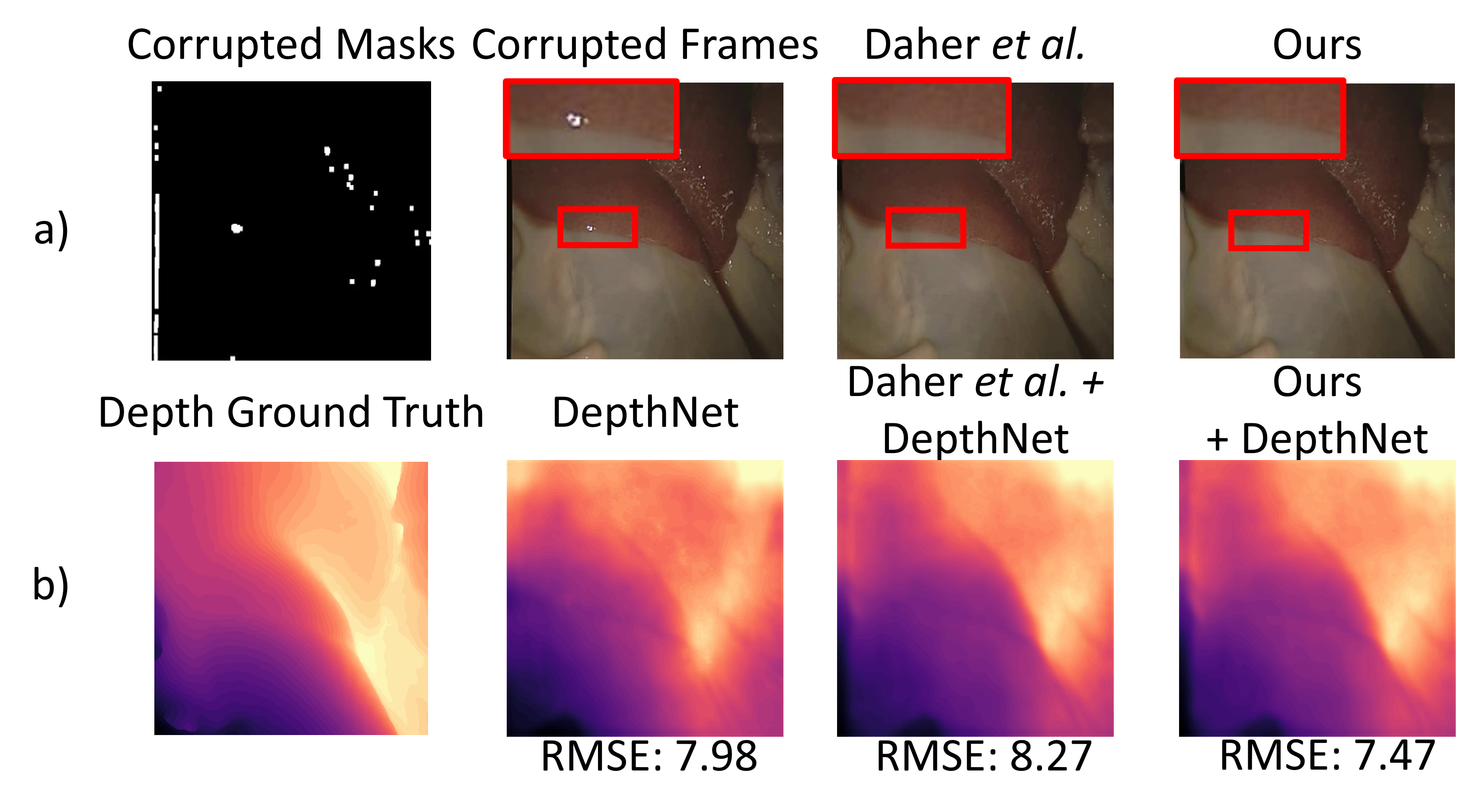}
    \caption{Comparison of deep learning-based inpainting performance on the SERV-CT dataset: a) Generalization Capability, and b) Depth Information Preservation.}
    \label{fig:dep}
\end{figure}
Additionally, Fig. \ref{fig:dep} a) shows our method's ability to inpaint highly plausible content on the SERV-CT dataset without specific fine-tuning, underscoring our approach's strong generalization capabilities. By applying the pre-trained DepthNet \cite{shao2022self} to our inpainted frames, Fig. \ref{fig:dep} b) reveals that our method improves depth estimation, achieving the lowest Root Mean Squared Error (RMSE) between the ground truth depth and the depth estimation after inpainting. This further underscores our method's superior effectiveness in preserving 3D spatial details compared to existing methods.

\noindent \textbf{Ablation Study}
Our ablation study in Table \ref{tab:SOTA} shows that our full framework yields the best results, highlighting the importance of each module. Notably, removing \ac{DED} lowers MSE and PSNR but doesn't affect SSIM, which remains comparable to the complete framework. This may be because \ac{DED} significantly enhances 3D spatial fidelity, which does not directly correspond to the features assessed by SSIM \cite{pambrun2015limitations}, such as contrast and luminance.
\begin{table}[t]
\scriptsize
\centering
\caption{Performance Analysis Across Depth Estimation Block Configurations.}
\begin{tabular}{l|p{1.9cm}|p{1.9cm}|p{1.9cm}}
\hline
Methods & $PSNR_{Crop} \uparrow$& $SSIM_{Crop}\uparrow$ &  $MSE_{Crop}\downarrow$ \\
\hline
DAEVI (First 4 Blocks) & \underline{29.921} & 0.792 &  \underline{100.851}\\
DAEVI (Last 4 Blocks) & 29.900 & \underline{0.796} &  101.313\\
\hline
DAEVI (All 8 Blocks) & \textbf{30.126} & \textbf{0.797} &  \textbf{97.873}\\
\hline
\end{tabular}
\label{tab:para}
\end{table}

\begin{table}[t]
\scriptsize
\centering
\caption{Online Inference Performance Analysis}
\begin{tabular}{l|p{2.3cm}|p{2.3cm}|p{2.3cm}}
\hline
Methods & $PSNR_{Crop} \uparrow$& $SSIM_{Crop}\uparrow$ &  $MSE_{Crop}\downarrow$ \\
\hline
Daher \emph{et al.} \cite{daher2023temporal} & 29.542 & 0.785 &  104.719\\
DAEVI & \textbf{30.126} & \textbf{0.797} &  \textbf{97.873}\\
\hline
DAEVI (Online) & \underline{30.117} & \textbf{0.797} &  \underline{98.147}\\
\hline 
\end{tabular}
\label{tab:online}
\end{table}

\noindent \textbf{Depth Estimation Block Configuration Analysis} 
Table \ref{tab:para} evaluates inpainting performance using different configurations of the first 4, last 4, and all 8 blocks of STGCN for depth estimation within the DAEVI framework. This analysis reveals that employing all 8 blocks, which combines both low-level and high-level features, achieves optimal performance, highlighting the effectiveness of our current STGDE design.

\noindent \textbf{Online Inference Performance Analysis}
Table \ref{tab:online} assesses our method's online inference ability, which employs only past frames for reference. Our online performance still outperforms the \ac{SOTA} method \cite{daher2023temporal} that uses both past and future frames as reference frames, demonstrating its potential for live endoscopic videos.

\section{Conclusion and Discussions}
In this paper, we propose the \ac{DAEVI} framework, the first endoscopic video inpainting framework designed to incorporate depth information to achieve reliable 3D spatial details. This work offers a potential solution to enhance the quality of endoscopic videos, facilitating informed clinical decision-making. Our \ac{DAEVI} leverages depth information from latent visual features of corrupted endoscopic frames with \ac{STGDE}, effectively fuses visual and depth information with \ac{BMPCF}, and assesses the fidelity of the RGB-D content with \ac{DED}. Experimental evaluations demonstrate its significant superiority compared to existing methods. 

While our framework demonstrates improvements in inpainting performance, it is not without limitations. As our work's focus is on corruption inpainting, the real-world applicability of \ac{DAEVI} could still be influenced by the effectiveness of any external corruption detection backbone. Future work could integrate existing endoscopic corruption detection methods such as semantic segmentation \cite{ali2021deep} into our framework, optimizing both corruption detection and inpainting tasks.
\begin{credits}
\subsubsection{\ackname}
This research is supported in part by the EPSRC NortHFutures project (ref: EP/X031012/1).
\subsubsection{\discintname}
The authors declare that there are no conflicts of interest related to this paper.
\end{credits}

\bibliographystyle{splncs04}
\bibliography{ref}

\end{document}


%
%
%
%
\title{Depth-Aware Endoscopic Video Inpainting -- Supplementary Material}
\authorrunning{Zhang et al.}    
\author{}
\institute{}
\maketitle  

\noindent \textbf{Reconstruction Loss} The details for $L_D$ and $L_I$ are as follows:
\begin{equation}
L_{D} = \left| \hat{D} - D \right|,
\end{equation}
\begin{equation}
L_{I} = \left| \hat{Y} - Y \right|,
\end{equation}
where $|\cdot|$ denotes the L1 Norm, $D$ and $\hat{D}$ denote the ground truth depth map and the translated depth map, respectively, and $Y$ and $\hat{Y}$ denote the ground truth frames and the inpainted frames, respectively.

\noindent \textbf{Perceptron and Style Loss} The details for $L_{P}$ and $L_{S}$ are as follows:
\begin{equation}
L_{P} = \sum_{l \in \text{Layers}} \frac{1}{N_l} \left| F_l(\hat{Y}) - F_l(Y) \right|_2^2,
\end{equation}
\begin{equation}
L_{S} = \sum_{l \in \text{Layers}} \left| G_l(\hat{Y}) - G_l(Y) \right|_F^2,
\end{equation}
where $F_l(\cdot)$ denotes the feature map extracted from layer $l$ of a pre-trained network given frames as input, and $G_l(\cdot)$ represents the Gram matrix of the feature map from layer $l$, capturing the style information. $|\cdot|_2$ denotes the squared Euclidean ($L2$) norm, and $|\cdot|_F$ denotes the squared Frobenius norm.

\begin{figure}[ht]
    \centering
    \includegraphics[scale = 0.38]{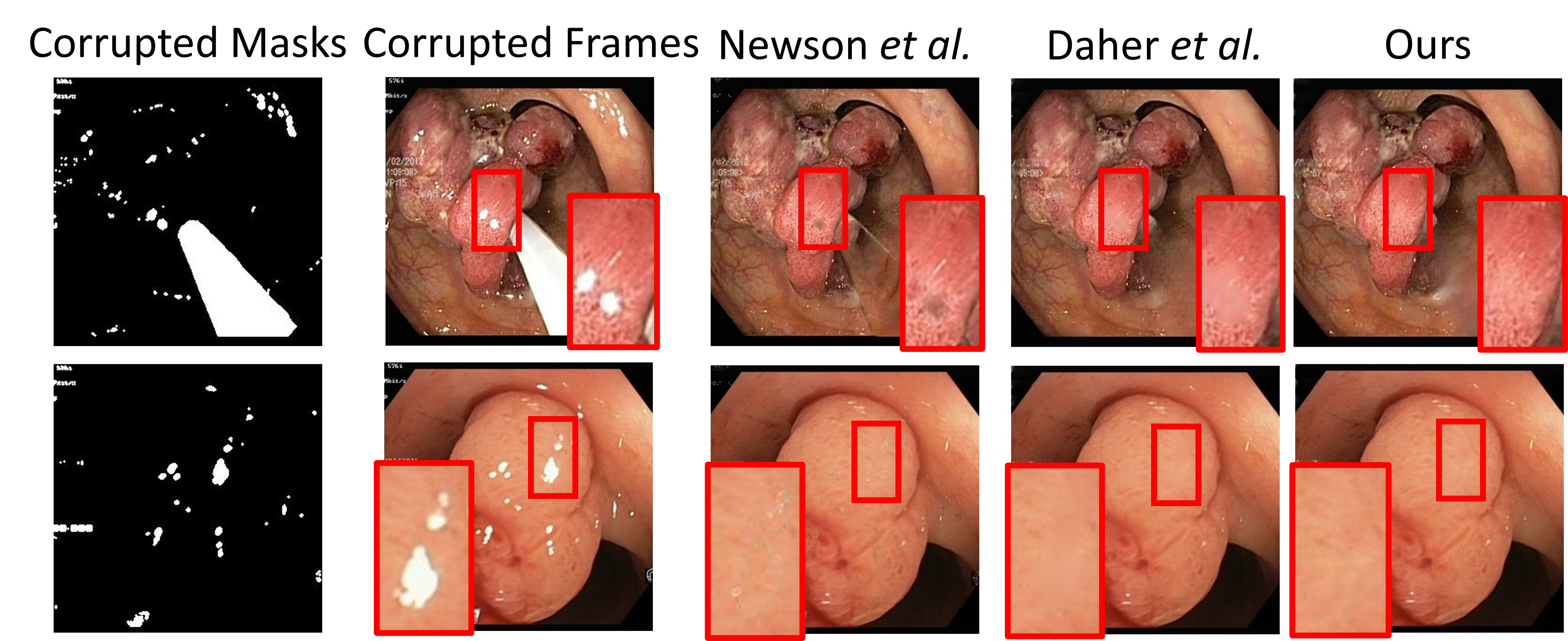}
    \caption{More Cases from the HyperKvasir Dataset: These cases further demonstrate that our method outperforms others, especially in generating fewer artifacts and more plausible details during endoscopic inpainting. This underscores our approach's superior corruption removal capability.}
    \label{fig:demo}
\end{figure}

\begin{figure}[ht]
    \centering
    \includegraphics[scale = 0.38]{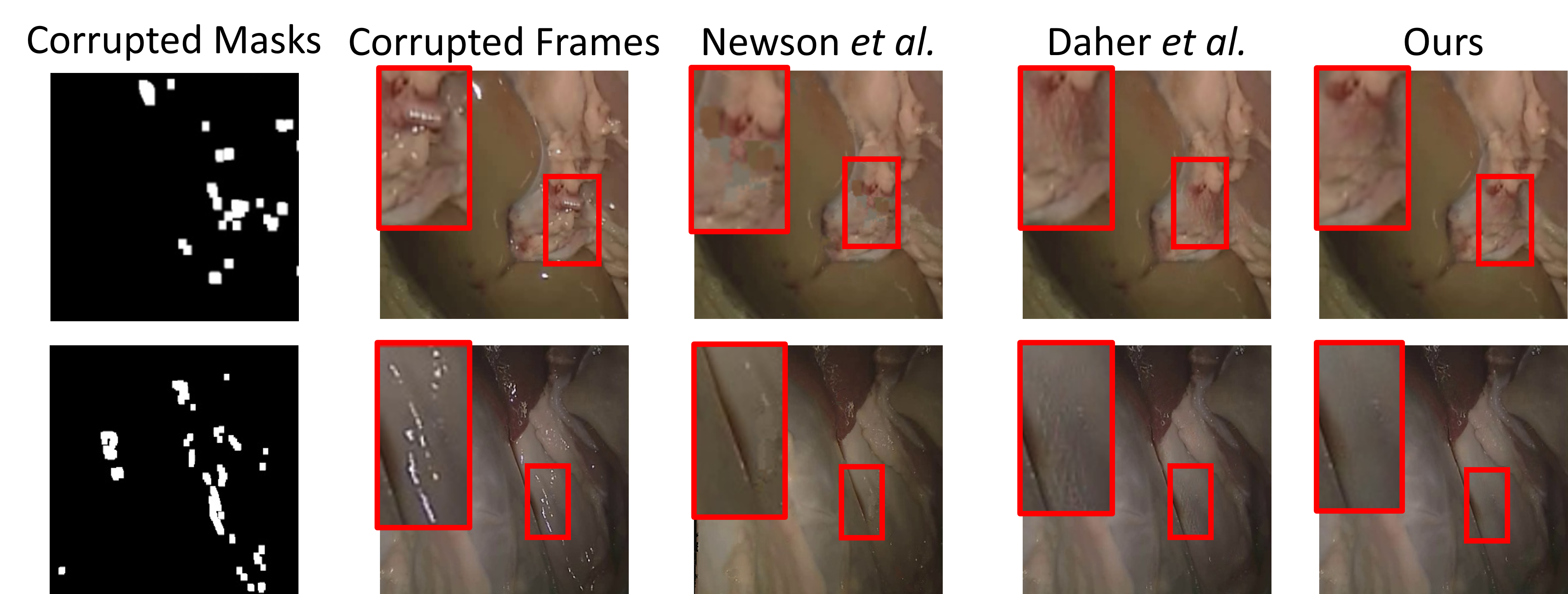}
    \caption{More Cases from the SERV-CT Dataset: These cases further demonstrate that our method outperforms others without the need for any fine-tuning, especially in generating fewer artifacts during inpainting. This underscores our approach's superior generalization capability.}
    \label{fig:demo}
\end{figure}

\begin{figure}[ht]
    \centering
    \includegraphics[scale = 0.38]{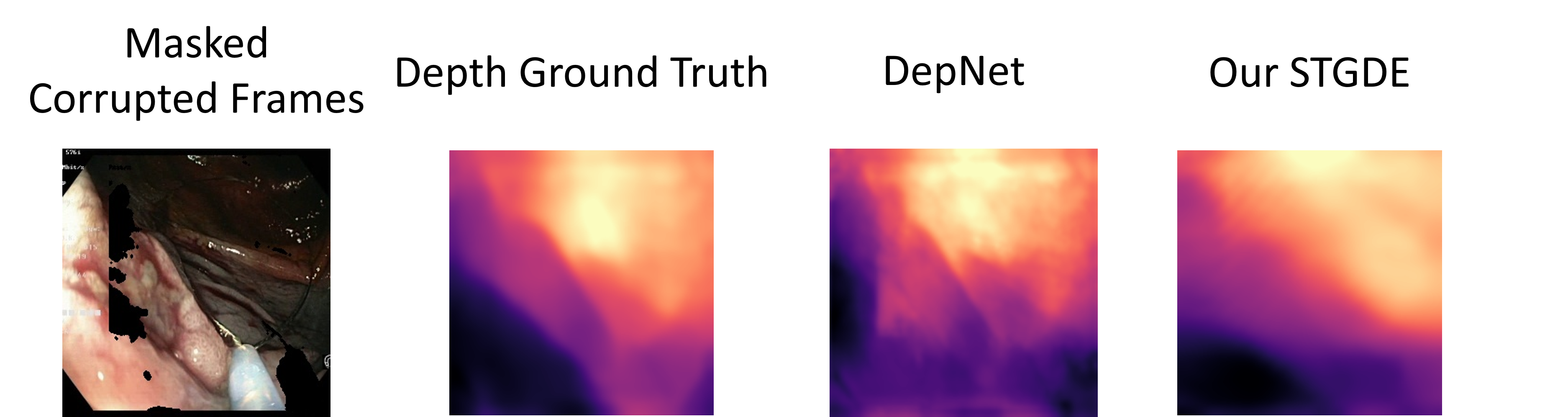}
   \caption{Depth Estimation Performance Analysis of Our Spatial-Temporal Guided Depth Estimation (STGDE) Module. This analysis compares the performance of our STGDE module against a pre-trained endoscopic depth estimator DepthNet, on masked corrupted frames. The ground truth is derived from depth estimation on unmasked frames. It is observed that our STGDE module estimates depth more accurately and closer to the ground truth compared to the direct application of the pre-trained model on masked frames.}
    \label{fig:demo}
\end{figure}

\newpage

\bibliographystyle{splncs04}
\bibliography{ref}